\documentclass[aps,preprint,prb,showpacs,amsmath,superscriptaddress,onecolumn,floatfix,nobalancelastpage]{revtex4-1}

\usepackage[nottoc,numbib]{tocbibind}

\usepackage[colorlinks=true, citecolor=black, linkcolor=black, urlcolor=black]{hyperref}

\usepackage[all]{hypcap}
\usepackage{amsmath}
\usepackage{enumerate,bm}
\usepackage{graphicx}
\usepackage{grffile} % to handle dots in graphics filnames
\usepackage{color}
\usepackage{esint}
\usepackage{textpos}
\usepackage[usenames,dvipsnames]{xcolor}
\usepackage{subfigure}
\usepackage{flushend}
\usepackage{tabularx}
\usepackage{amssymb}
\usepackage{float}
\usepackage{subfigure}
\usepackage{ulem} %striketrhough

\usepackage{setspace}

\begin{document}
% \linenumbers

\author{Hannes~H\"ubener}
\email{hannes.huebener@gmail.com}
\affiliation{Max Planck Institute for the Structure and Dynamics of Matter and Center for Free Electron Laser Science, 22761 Hamburg, Germany}

\author{Umberto~De~Giovannini}
\email{umberto.degiovannini@gmail.com}
\affiliation{Max Planck Institute for the Structure and Dynamics of Matter and Center for Free Electron Laser Science, 22761 Hamburg, Germany}

\author{Angel~Rubio}
\email{angel.rubio@mpsd.mpg.de}
\affiliation{Max Planck Institute for the Structure and Dynamics of Matter and Center for Free Electron Laser Science, 22761 Hamburg, Germany}

%\title{Dynamical dressing of electronic quasiparticles by coherent bosonic fields}
%\title{Dynamical dressing of electronic quasiparticles by coherent-phonons}
\title{Phonon driven Floquet matter}
\date{\today}

\begin{abstract}
The effect of electron-phonon coupling in materials can be interpreted as a dressing of the electronic structure by the lattice vibration, leading to vibrational replicas and hybridization of electronic states. In solids a resonantly excited coherent phonon leads to a periodic oscillation of the atomic lattice in a crystal structure bringing the material into a non-equilibrium electronic configuration. Periodically oscillating quantum systems can be understood in terms of Floquet theory, which has a long tradition in the study of semi-classical light-matter interaction. Here, we show that the concepts of Floquet analysis can be applied to coherent lattice vibrations reflecting the underlying coupling mechanism between electrons and coherent bosonic modes. This coupling leads to phonon- or photon-dressed quasi-particles imprinting specific signatures in the spectrum of the electronic structure. Such dressed electronic states can be detected by time- and angular-resolved photoelectron spectroscopy (ARPES) manifesting as sidebands to the equilibrium band structure. Taking graphene as a paradigmatic material with strong electron-phonon interaction and non-trivial topology we show how the phonon-dressed states display an intricate sideband structure revealing the electron-phonon coupling at the Brillouin zone center and topological ordering of the Dirac bands. Most strikingly, we find that the non-equilibrium electronic structure created by coherent dynamical dressing is the same for photon and phonon perturbations. We demonstrate that if time-reversal symmetry is broken by the coherent lattice perturbations a topological phase transition can be induced. This work establishes that the recently demonstrated concept of light-induced non-equilibrium Floquet phases can also be applied when using coherent phonon modes for the dynamical control of material properties. The present results are generic for bosonic time-dependent perturbations, therefore we envision similar phenomena to be observed for example for plasmon, magnon or exciton driven materials.
\end{abstract}

\maketitle

\section{Introduction}
Interaction of electrons and bosons in solids and molecules often leads to satellite features in the electronic structure resulting from harmonics of the boson mode. A well known example are plasma oscillations in solids that are detectable as satellites in the photoelectron spectrum.\cite{Lischner:2015fd,Caruso:2015fj} Another one is the effect of vibrational motion of ions that provides an intrinsic mode of the material that is detectable in the electronic structure, most prominently as vibrational sidebands in optical absorption spectroscopy of molecules.\cite{Holland:1997bi} This is facilitated by the coupling of the electrons to the vibrational modes, also known as phonons in crystals. If the ions perform a collective oscillation, then the quantum mechanical wavefunctions of the ions are coherent superpositions of harmonic oscillator states forming wavepackets that move in analogy to the trajectory of a classical particle.\cite{Kuznetsov:1994jv} In such a state the material effectively has a periodically oscillating lattice, which means for the electrons of the system that they are experiencing a time-dependent ionic potential. The question we are answering here is: what is the effect of such an oscillation on the electronic structure of a material and how can one define, observe and understand such a non-equilibrium phase? 

In equilibrium a strong electron-phonon coupling can result in a dressed electronic structure where the dressing creates observable replica bands\cite{Lee:2014bw}  or kinks\cite{Graf:2008bs,Siegel:2012bc,Mazzola:2013if} in the electronic spectrum. In such a strongly coupled material a few phonon quanta have a large effect on the electrons, whereas in a weaker coupled system one needs more quanta to achieve a similar effect. Creating more quanta of a boson mode means that one has to excite it coherently, thereby driving the system out of its equilibrium. On a general note, out of equilibrium materials have a rich variety of properties that can be fundamentally different from their equilibrium states. Often such properties occur only transiently which prevents further investigation or technological applications. It is therefore of great interest to create non-equilibrium states that can be stabilized, controlled and manipulated to unravel these hidden material phases. 

The controlled excitation of lattice vibrations as coherent phonon modes\cite{Forst:2011ep} has been shown to provide an avenue towards engineering long-lived, transient non-equilibrium states, such as light-enhanced\cite{Hu:2014cg} and light-induced superconductivity,\cite{Mitrano:2016fr} vibrationally controlled\cite{Rini:2007hc,Forst:2015fv} and induced\cite{Nova:2017ja} magnetism and phonon control of ferro-electricity\cite{Mankowsky:2017ur} among others. This kind of non-equilibrium boson driven phases might also be used to affect exciton\cite{Kasprzak:2006jy} and polariton-condensates\cite{CerdaMendez:2010bf} and Higgs-modes superconductors.\cite{Matsunaga:2014kj,Sherman:2015eh} In particular, the proposed ability to induce a superconducting phase in materials that are ordinary conductors in equilibrium by optically exciting coherent phonons,\cite{Forst:2014eq,Kaiser:2014je,Mankowsky:2014em,Mitrano:2016fr} has spurred theoretical investigations into the electronic properties of materials under such non-equilibrium conditions.\cite{Kennes:2017fu,Sentef:2017,Babadi:2017}
%our spin-floquet paper:
%Recently, the possibility to affect spin-polarization and magnetization through phonon-driving has been proposed for transition metal dichalcogenides.\cite{Shin:2017} 

%Excitation of a material with a laser that has the energy corresponding to a phonon frequency can lead to the creation of a coherent phonon, that is the collective motion of all lattice ions with the frequency corresponding to the energy of the phonon. The quantum mechanical wavefunctions of the ions are in this case coherent superpositions of harmonic oscillator states forming wavepackets that move in analogy to the trajectory of a classical particle.\cite{Kuznetsov:1994jv} In such a state the material effectively has a periodically oscillating lattice, which means for the electrons of the system that they are experiencing a time-dependent ionic potential. The question we are answering here is: what is the effect of such an oscillation on the electronic structure of a material and how can one define, observe and understand such a non-equilibrium phase? 

We show how the electronic structure together with the time-periodic ionic potential created by a coherent phonon mode can be interpreted as a steady state, quasi-static system when observed for times longer than the phonon time-scale. Such a Floquet state where the coherent phonon dresses the electronic states represents a distinct non-equilibrium phase with a fundamentally altered electronic structure. This kind of non-equilibrium steady state picture of a dressed electronic structure has a long tradition when considering light-matter interaction, going back all the way to the idea of electrons being dressed by photons in the optical Stark effect.\cite{Autler:1955gb,Shirley:1965cy,Fainshtein:1978bs,Grifoni:1998bm,Sie:2015hn,deGiovannini:2016cb} However, recently there has been considerable activity in the theoretical framework of Floquet-theory for light-driven matter following the proposal of the Floquet topological insulator.\cite{Lindner:2011ip,Sentef:2015jp,Claassen:2016ge,Fleury:2016cr,Zou:2016do,Hubener:2017ht} The Floquet-phonon framework proposed here can be used as a paradigm to treat vibrationally excited systems and is thus for example also applicable for shaken optical lattices, where instead of a phonon mode a macroscopic vibration is applied.\cite{Choi:2017ho}  It is equally applicable to the interpretation of vibrational spectroscopy in molecules, where the phonon-dressed sidebands occur as a series of vibrational peaks. Moreover, we posit that this framework is transferable to other coherent bosonic excitations that can be expressed as semi-classical fields such as plasmons or magnons. 

In contrast to the interaction of optical light with electrons, where the massless photons dress the electrons with fast oscillating fields, the picture is reversed when considering the effect of a coherent phonon on the electronic structure. The relatively slow movement of the much heavier ions, allows for the electrons to follow this movement almost instantaneously. Indeed, this picture gives rise to the frozen phonon approximation, where one assumes that the electrons experience any given ionic configuration of the lattice as if the lattice was fixed. While this approximation has been used successfully in a wide range of applications,\cite{Harmon:1981it,Kunc:1981bq,Weyrich:2011bh,VanDyck:2009is,Budai:2014dl,Fausti:2014ts,PhysRevB.92.035147,Rossi2015} we show that it fails to accurately represent the electronic structure under the Floquet-phonon conditions. The frozen phonon approximation cannot account for the dynamical dressing effect that the coherently oscillating lattice supplies to the electronic structure. By using a first-principles Floquet description we show that one can define observable band structures of the dynamical system that are fundamentally different from the frozen phonon approach. 
%HH 
The failure of the frozen phonon approximation to correctly describe dynamical effects of the electron-phonon coupling in graphene has been pointed out before\cite{Pisana:2007cl}. The lattice vibration dynamically changes the electronic structure, which in turn leads to a large renormalization of phonon modes. The Floquet-phonon approach presented here can serve as a starting point to compute such a dynamical renormalization of phonon energies, resulting in a non-equilibrium phonon bandstructure.

We present results obtained from two different complementary types of first-principles calculations, by first computing the real-time propagation of a quantum mechanical electronic system together with classical movement of the ions. In a second approach we analyze the periodic oscillation induced by this motion with Floquet theory, thereby obtaining a quasi-static representation of the dynamics which allows us to explain in detail spectroscopic features of the time-evolution and analyze them in terms of the underlying multi-phonon band structure. In this work we are considering the in-plane double degenerate optical phonon of graphene as a paradigm for a coherently driven phonon system. While these modes are not infrared active in monolayer graphene, they can become active in bilayer graphene.\cite{Gierz:2015gn} The features we discuss are generic for any semimetal and thus have readily transferable implications for driven charge-density wave and electron-phonon superconductors. The topology of the Dirac bands on the other hand is specific to topological systems and our results represent a general dynamically induced topological phase transition. 

\begin{figure}
   \resizebox{0.9\textwidth}{!}{\includegraphics{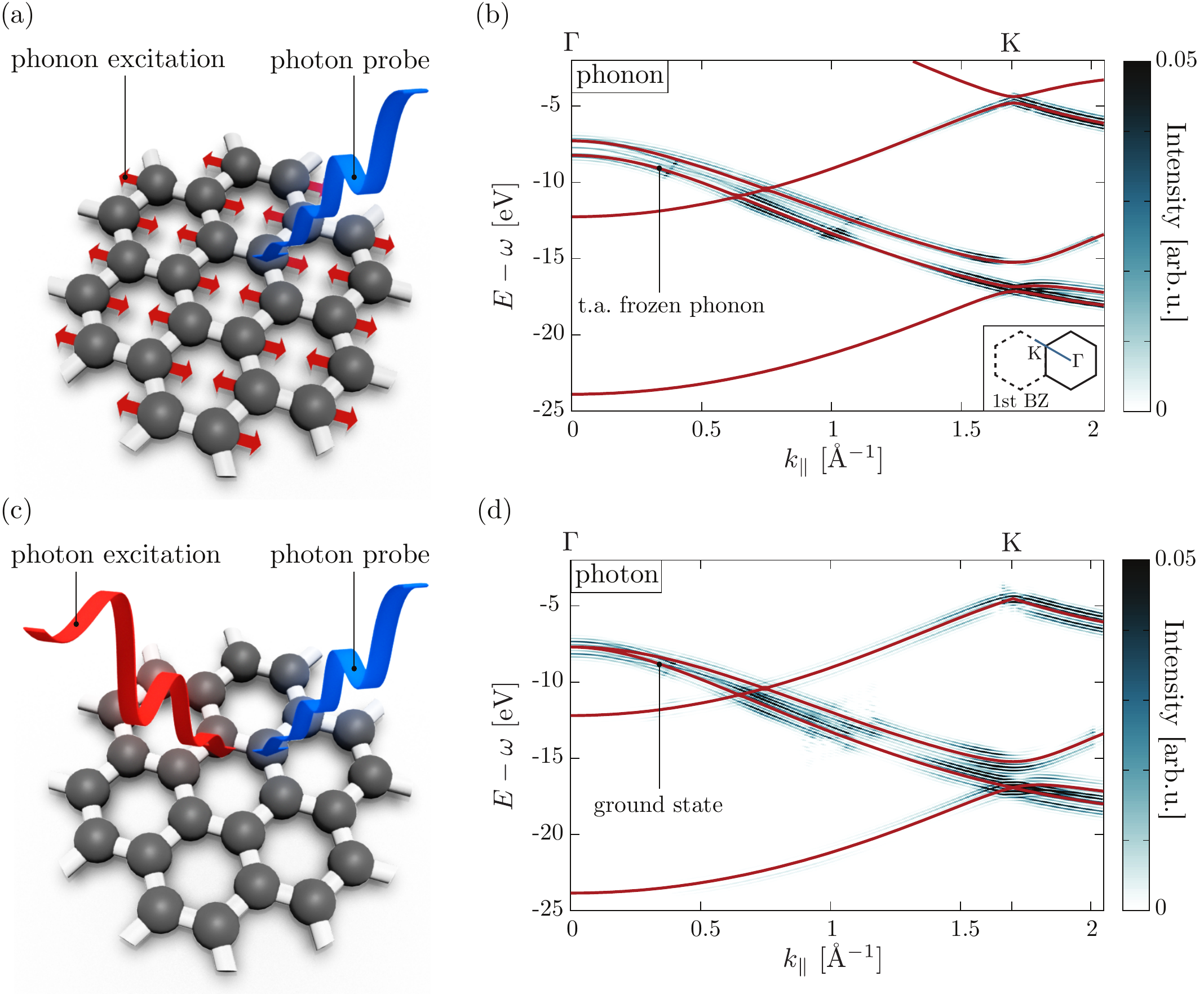}}
   \caption{Time-resolved ARPES of graphene under different fundamental excitations: (a) Sketch of the $E_{2g}$ optical mode in graphene that has been prepared as a coherent lattice motion while the photo-electron spectrum (b) is computed. (b) The computed time-resolved ARPES spectrum of graphene under driving by a coherent phonon. In red is shown the frozen phonon band structure averaged over a phonon cycle (see text). (c) Sketch of an optical excitation of the electronic structure with fixed ions as used for the time-resolved ARPES calculation in (d).  (d) Under only optical excitations the time-resolved ARPES specturm displays a similar sideband splitting as the phonon-dressed system  throughout the Brillouin zone, but the strongly interacting pattern at the $\Gamma$-point is absent. In red is shown the groundstate bandstructure.
   %HH mod
   The inset to (b) shows the path in momentum space that is chosen to cross the second Brillouin zone, because it provides a stronger photo-electron signal.
   }
\end{figure}

\section{Results}
To observe the effect of a coherent lattice motion on the electronic structure of graphene we calculate the time- and angular-resolved photoelectron spectroscopy (ARPES) spectrum, by using time-dependent density functional theory (TDDFT)\cite{runge_density-functional_1984,bertsch_real-space_2000} coupled to Ehrenfest molecular dynamics for the classical motion of the ions according to the optical $E_{2g}$ mode of graphene, c.f. Fig.~1 and Methods for technical details. We assume that after an initial excitation of the coherent phonon the pump laser is switched off, but the phonon mode maintains its coherence for some time in which the system is probed. In our simulations we use probe times of up to 160~fs which is well below reported coherence times for phonon modes of about 1~ps.\cite{Jeong:2015tn} The resulting photo-emission spectrum for a probe pulse longer than the $E_{2g}$ phonon period of 20.6~fs, corresponding to the frequency of this mode at the $\Gamma$-point of the phonon-Brillouin zone, is shown in Fig.~1(b). We show photoelectron spectra for the second Brilluoin zone throughout this paper, because they display a stronger photo-electron signal. For comparison with the first Brillouion zone, see supplementary Fig.~S3. The frozen phonon picture assumes that at any given time the electrons are in the equilibrium corresponding to the lattice configuration at that time. This suggests, as an \textit{ad hoc} approximation for the time-resolved ARPES spectrum, to take the average the different frozen-phonon band structures over the time of measurement. Such a time-averaged frozen-phonon band structure is shown for comparison with the time-resolved ARPES result in Fig.~1(b). By contrast, we show in Fig.~1(d) the corresponding time-resolved ARPES spectrum where only an optical pump pulse is used and the ions are kept fixed in their equilibrium positions so that only the effect of photo-dressing is observed. The different behaviour of the electronic structure is most striking at the $\Gamma$-point of the Brillouin zone. Overall the two different dressing mechanisms result in a similar structure that is characterized by an equally spaced stacking of sidebands around an equilibrium band. The degenerate top valence band at $\Gamma$, known as the $\sigma$-bands, however, display a qualitatively different behaviour as becomes clear in Fig.~2(a): The photoelectron spectrum shows an apparent split electronic bands around the degenerate groundstate bands which seems to be in agreement with the cycle averaged frozen phonon band structure. This agreement, however, is deceptive and the underlying mechanism is more involved, as will become clear below, by considering the Floquet spectrum. Above all, we point out that a frozen phonon description lifts the degeneracy of the top valence band, which is clearly not observed in the photoelectron spectrum, where the position of the groundstate bands is preserved. While, in these calculations we consider only the specific lattice vibration associated with the coherent $E_{2g}$ phonon, we have also performed calculations with an additional random (thermal) distribution of ionic motion and find the same photo-electron spectrum (c.f Supplementary Fig.~S1). 

\begin{figure}
   \resizebox{0.8\textwidth}{!}{\includegraphics{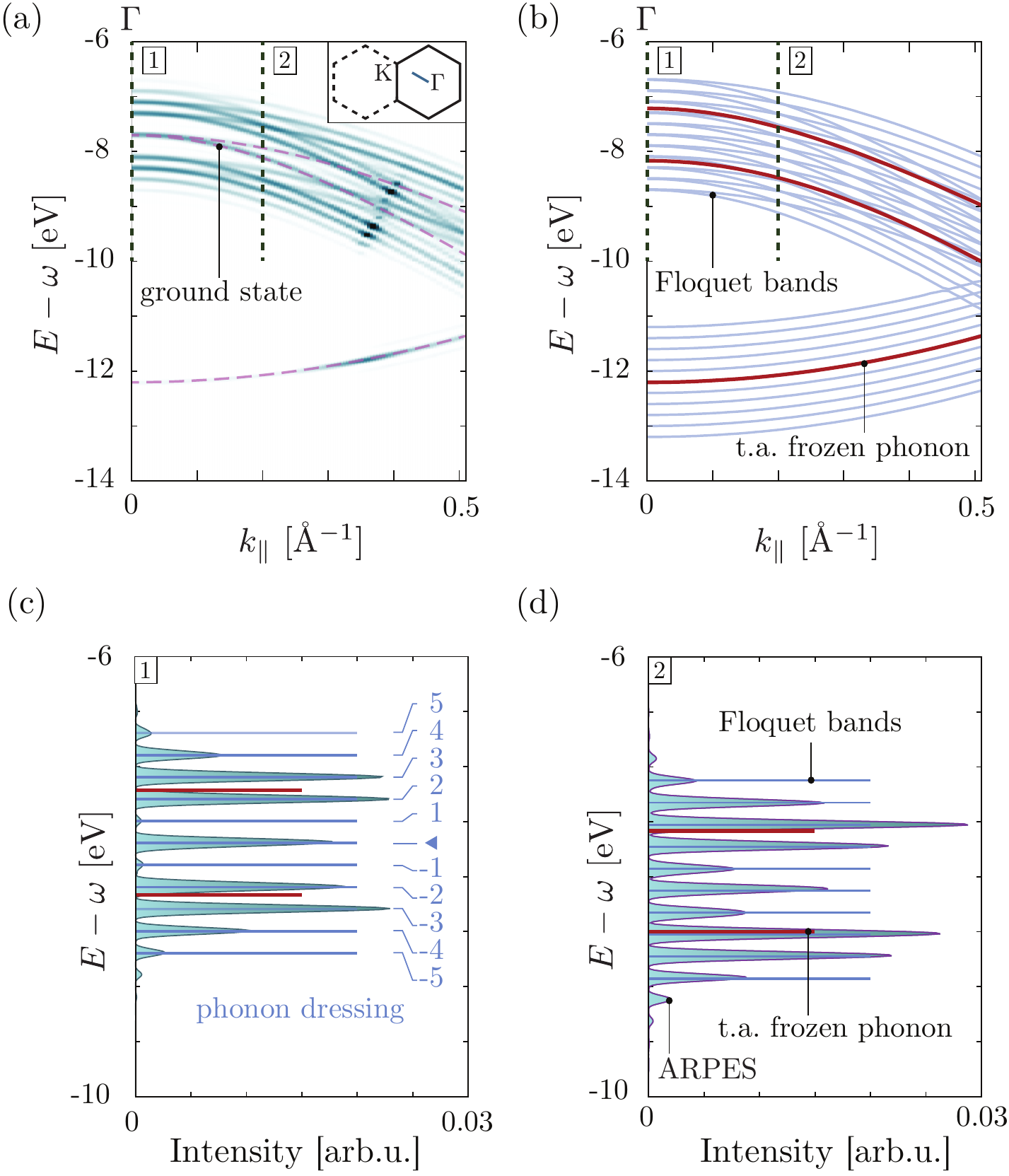}}
   \caption{Time-resolved ARPES of graphene at the $\Gamma$ point with a coherent phonon: (a) Computed time-resolved ARPES spectrum of  for a small path in the $\Gamma$-$K$ direction of the Brillouin zone (see inset). An intricate sideband structure originating from the coherent phonon excitation is visible. (b) Band structure computed from Floquet-analysis of the first-principles time-dependent Hamiltonian.  In red is shown the time averaged frozen phonon band structure (see text) that can be seen failing to give the essential spectral features. (c) and (d) show vertical cuts through the time-resolved ARPES spectrum shown as dashed lines in (a) together with the Floquet-energy levels taken at the same position, as indicated by the dashed lines in (b). In (c) the Floquet bands are indexed according to their phonon-multiplicity and the position of the original band is indicated by a triangle. Both cuts show the excellent agreement between the Floquet bands and the time-resolved ARPES spectrum. The low intensity of the first sidebands of the $\sigma$ bands at the $\Gamma$-point is due to weak photo-electron matrix elements. The cycle averaged frozen phonon bands (red) do not coincide either with Floquet-levels nor with peaks in the calculated time-resolved ARPES spectrum.}
\end{figure}

While the cycle averaged frozen-phonon bands seemingly capture the redistribution of the photo-electron spectral weights qualitatively correctly, they do not contain any information about the side-band structure and fail to account for the central band in the photo-electron spectrum. To obtain the electronic structure underlying the bands observed in the photoemission spectrum we use Floquet analysis of the time-dependent Hamiltonian generated by the TDDFT calculation.\cite{Hubener:2017ht} In this method a stationary state is expanded into a basis of Fourier components of multiples of the mode frequency $\Omega$: $|\Psi_\alpha(t)\rangle=\sum_{m}\exp(-i(\epsilon_\alpha+m\Omega)t)|u_m\rangle$, where $\epsilon_\alpha$ is the quasi-static Floquet band. With this ansatz the time-dependent Schr\"odinger equation becomes an eigenvalue problem $\sum_{n}\mathcal{H}^{mn} |u_n\rangle = \epsilon_\alpha |u_m\rangle$ of the static Floquet Hamiltonian $\mathcal{H}^{mn}  = \frac{\Omega}{2 \pi}\int_{2\pi/\Omega} dt e^{i(m-n)\Omega t} H(t) + \delta_{mn}m\Omega$. The eigenstates of this Hamiltonian span a Hilbert space with the dimension of the original electronic Hilbert space times the multi-mode (photon or phonon) dimension. The contribution of the latter is in principle infinite, but can be truncated to a number large enough to capture the interaction between sidebands (see for instance Supplementary Fig.~S2). The spectrum of this Hamiltonian gives the band structure of the dressed quasiparticles. 

Floquet theory provides the correct way of performing the quantum-mechanical time-average over a cycle period of a steady state oscillating system and thus represents the finite time duration of the ARPES probe.\cite{Sentef:2015jp,DeGiovannini:2016bb} We note that in order to include the time-dependent Hamiltonian of a system with moving ions into the Floquet integrals it is essential to use a generic representation, such as real-space grids or plane-waves. A local basis, like atomic centered orbitals, or the Kohn-Sham eigenstate representation of the Hamiltonian cannot be used here, because it precludes the possibility to perform an integral over the Hamiltonian at different times. We also note, that because of the low energy of the dressing field, we cannot use a high-frequency expansion that is otherwise convenient for analytic treatments. The results of the Floquet-TDDFT calculation of the coherent-phonon excited electronic system is shown in Fig.~2(b) and the excellent agreement between the Floquet bands and the time-resolved ARPES spectrum of Fig.~2(a) is apparent.

The sidebands of the $\sigma$-bands does not follow an underlying splitting of the bands, contrary to what the \textit{ad hoc} frozen phonon approximation suggests. Instead of a splitting of the bands, the degeneracy is preserved even for the sidebands and comparison with the Floquet spectrum reveals that the first two sidebands are surpressed by photo-electron matrix elements. In Fig.~2(c) and (d) the time-resolved ARPES intensity for these bands around $\Gamma$ is given in more detail, showing the excellent quantitative agreement between Floquet theory and the time-resolved ARPES spectrum, as well the failure of the frozen-phonon approximation to quantitatively reproduce even the splitting of the spectral weight.

The necessity of Floquet-theory to correctly describe the dynamical dressing becomes even clearer when looking at the $K$ point of the Brillouin zone. In graphene the Dirac point at $K$ is of particular interest because here the electronic structure can be represented by the two-level Dirac Hamiltonian, reflecting the particular topological properties of the material. Figs.~3~and~4 show an enlarged view of differently (phonon and photon) dressed electronic structures around $K$. By comparing the Floquet band structure of the $E_{2g}$ mode, Fig.~3(a), with the one of a linearly polarized photon field, Fig.~3(c), we find that by choosing the appropriate amplitudes one can obtain identical band structures for both types of excitation. This remarkable agreement is a consequence of the same nature of the underlying electron-boson coupling. The insets in Figs.~3(a) and (c) show that neither the photon nor the phonon perturbation leads to the creation of a bandgap at the Dirac point.

The analogy of photon and phonon dressing at the $K$ point can be further exploited by considering a coherent phonon excitation that is a linear combination of the two degenerate longitudinal (LO) and  transverse (TO) modes of $E_{2g}$ to create a circular, time-reversal symmetry breaking motion in analogy to a circularly polarized electromagnetic (photon) field. Indeed, as shown in Figs.~4(a) and (c) both circular excitations result again in identical band structures. In this case, as shown by the insets of Figs.~4(a) and (c), the perturbation leads to the creation of a dynamical gap. Such an excitation is described by the Haldane model\cite{haldane_model_1988} and can be associated with a topological phase transition. This dynamical opening of the gap is a general feature of a Floquet-topological phase\cite{Sentef:2015jp} in graphene which, as will become clear below, is also a property of the coherent lattice excited state. Here, we point out that the apparent similarity of the non-equilibrium electronic structure at the $K$ point of graphene for different bosonic excitations is a consequence the same underlying coupling mechanism of the Dirac bands. %Furthermore, we observe a qualitative failure of the time-averaged frozen-phonon approach, in Figs.~3(a) and (b). For both phonon polarizations the frozen phonon approach gives a gapped band structure, while only in the case of circular excitation the Floquet bands structure has a finite band gap.  

%\begin{figure}
%   \resizebox{0.9\textwidth}{!}{\includegraphics{fig3.pdf}}
%   \caption{Time-resolved ARPES of graphene at the $K$ point: (a) The coherent phonon motion depicted in the sketch leads to a time-resolved ARPES spectrum that is very well described by the corresponding TDDFT-Floquet band structures. In particular the degeneracy of the Dirac bands at the Dirac point can be seen to be preserved by Floquet theory (see inset), while the corresponding cycle-averaged frozen phonon band (in red) fails. (b) The excitation of a circularly polarized coherent phonon breaks time-reversal symmetry and leads to a qualitatively different electronic spectrum. In this case the degeneracy of the Dirac points is lifted (see inset) and the system is in a non-trivial topological phase. (c) and (d) show the corresponding time-resolved ARPES spectra where the ions are fixed and only the effect of photon-dressing is shown for a linear (c) and circular (d) polarized field. While the time-resolved ARPES show different intensity patterns than for the corresponding phonon-dressed cases, the Floquet-band structures reveal that for the Dirac point the phonon- and the photon-dressing create the same kind of non-equilibrium phase. The dynamical effect of the lattice perturbation on the electronic structure can be described by an effective gauge field, indicated by $A^u$, that plays the same role as the physical electro-magnetic vector potential $A^\gamma$ of the photon field.} 
 %  (A=6 circular A=4 linear polarized along $x$.)
%\end{figure}

\begin{figure}
   \resizebox{0.9\textwidth}{!}{\includegraphics{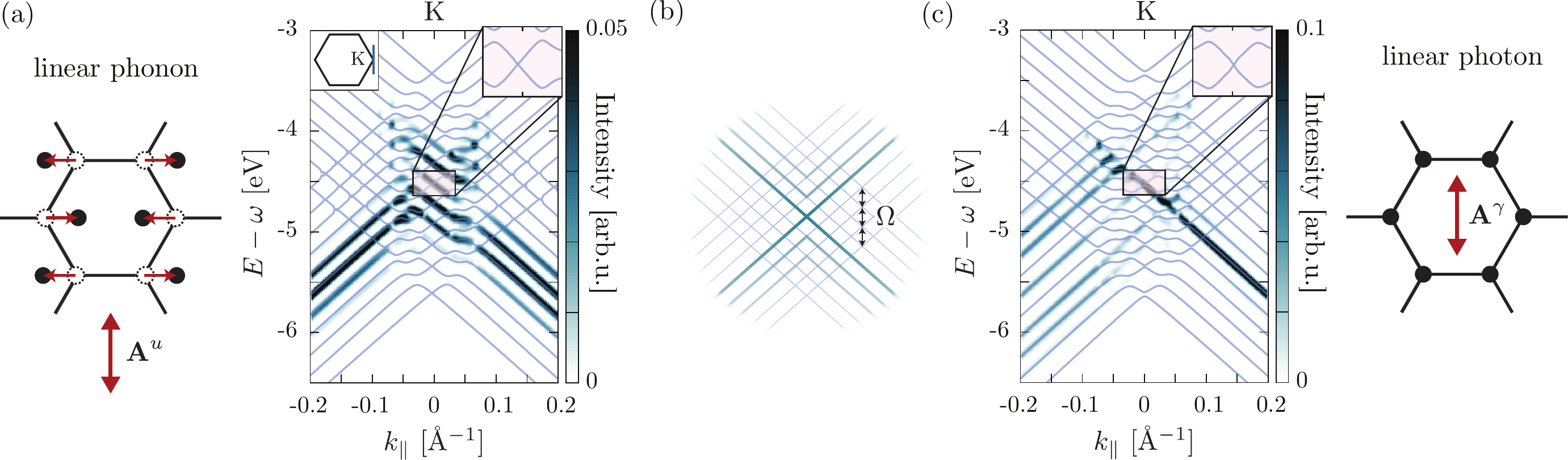}}
   \caption{Time-resolved ARPES of graphene at the $K$ point with linearly polarized pumping: (a) The coherent phonon motion depicted in the sketch leads to a time-resolved ARPES spectrum that is very well described by the corresponding TDDFT-Floquet band structures. In particular the degeneracy of the Dirac bands at the Dirac point can be seen to be preserved by Floquet theory (see inset). By contrast in (c) is shown the ARPES spectrum of a corresponding electronic structure where the ions are fixed and only the effect of a photon pump is included. Except for differences in the the photo-electron intensities, the two dressing mechanism result in the same non-equilibrium electronic structure. The dynamical effect of the lattice perturbation on the electronic structure can be described by an effective gauge field, indicated by $A^u$, that plays the same role as the physical electro-magnetic vector potential $A^\gamma$ of the photon field. Panel (b) indicates how to identify the sidebands in the band diagrams, as originating from a regular mesh of shited Dirac bands.} 
 %  (A=6 circular A=4 linear polarized along $x$.)
\end{figure}

\begin{figure}
   \resizebox{0.9\textwidth}{!}{\includegraphics{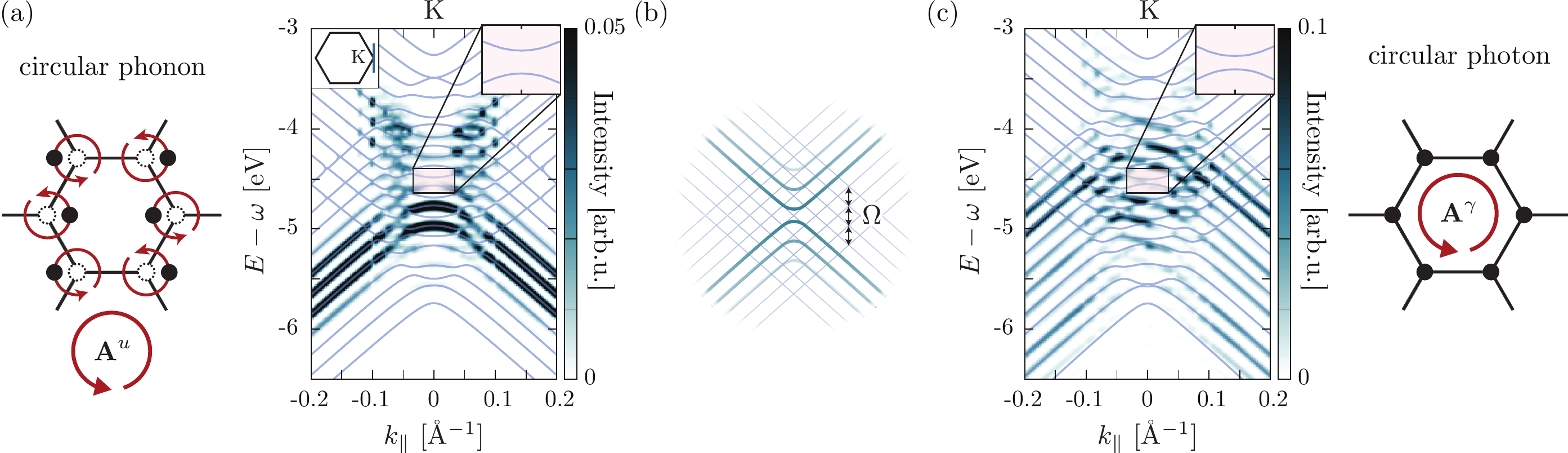}}
   \caption{Time-resolved ARPES of graphene at the $K$ point with circularly polarized pumping: (a) The coherent phonon motion depicted in the sketch leads to a time-resolved ARPES spectrum that is very well described by the corresponding TDDFT-Floquet band structures. In particular the  degeneracy of the Dirac points is lifted (see inset) and the system is in a non-trivial topological phase. (c) shows the corresponding time-resolved ARPES spectra where the ions are fixed and only the effect of photon-dressing is shown. While the time-resolved ARPES show different intensity patterns than for the corresponding phonon-dressed case, the Floquet-band structures reveal that for the Dirac point the phonon- and the photon-dressing create the same kind of non-equilibrium phase. The dynamical effect of the lattice perturbation on the electronic structure can be described by an effective gauge field, indicated by $A^u$, that plays the same role as the physical electro-magnetic vector potential $A^\gamma$ of the photon field. Panel (b) indicates how to identify the sidebands in the band diagrams, as originating from a regular mesh of shited Dirac bands and the induced opening of a gap at band crossing points.} 
 %  (A=6 circular A=4 linear polarized along $x$.)
\end{figure}

\section{Discussion}
Using Floquet theory the similarities of the dressed electronic structure at the Dirac point can be traced back to the coupling between the electrons and the coherent boson fields. The coupling for the phonon case is described by electron-phonon matrix elements\cite{Giustino:2017ge} of the $E_{2g}$ mode, while for the photon case is mediated by dipole matrix elements with polarization corresponding to the pump laser. To first order the coupling of electrons to a coherent boson field is given by the Hamiltonian $H(t) = \sum_i \epsilon_i c^\dagger_i c_i +  \sum_i m_{i,j}  c^\dagger_i c_j\hat{\mathbf{F}}(t)$
where $c^\dagger$ and $c$ are the fermionic creation and annihilation operators, $\epsilon$ are the energies of the uncoupled electrons and $m$ are the electron-boson coupling matrix elements. $\hat{\mathbf{F}} = a^\dagger + a$ is the coherent boson field, which corresponds to the classical limit of the quantized field, oscillating with the frequency $\Omega$. In the case of phonons this is the time-dependent lattice displacement, $\hat{\mathbf{F}}(t) = \mathbf{u} \sin(\Omega t)$ along oscillation direction $\mathbf{u}$ of the phonon, while for photons it is the classical vector potential of the laser polarized in $\mathbf{A}^{\gamma}$ direction, $\hat{\mathbf{F}}(t) = \mathbf{A}^{\gamma} \sin(\Omega t)$.

For the Dirac bands around $K$ the groundstate Hamiltonian can be written as a two level system $H^D_0=v_F (k_x \sigma_x+k_y \sigma_y)$, where $\sigma_i$ are Pauli matrices and $k_i$ the in-plane components of the crystal momentum centered at the Dirac point.\cite{haldane_model_1988} Expanding the time-dependent (boson coupled) Hamiltonian in the eigenbasis of the groundstate Hamiltonian yields (c.f Supplementary for details)

\begin{equation}
 H^D_\mathbf{k} (t) =v_F |\mathbf{k}|\sigma_z + m [\cos(\theta_\mathbf{k}) \sigma_z  + \sin(\theta_\mathbf{k})\sigma_y] \sin(\Omega t) .
\end{equation}

The angle $\theta_\mathbf{k}$ is the angle between the polarization of the vector potential $\mathbf{A}^{\gamma}$ and the $\mathbf{k}$-vector for the electromagnetic (photon) excitation. For the phonon case instead $\theta_\mathbf{k}$ is the angle between $\mathbf{k}$ and a vector $\mathbf{A}^{u}$ that is perpendicular to the phonon-polarization, i.e. $\mathbf{A}^{u}\cdot \mathbf{u} = 0$. The Hamiltonian Eq.~(1) is a particular representation of the Peierl's substitution describing weak coupling to a classical field, i.e.  where $\mathbf{k}\rightarrow \mathbf{k}-\mathbf{A}^i$. This becomes evident when considering the eigenvalues of this operator for fixed time: $E^\pm_{[t]} = \pm v_F |\mathbf{k}-\mathbf{A}^i_{[t]}|$. Here we have parametrised the time-dependence to emphasize that such instantaneous eigenvalues do not have any physical meaning for a vector potential. Nevertheless, it shows the effect of the dressing field, either photonic or phononic, on the electronic Hamiltonian as an effective gauge field in minimal coupling (Peierl's substitution) form. In particular, it shows that the lattice deformation induced by the phonon displacement $\mathbf{u}$ can be described by an effective gauge field $\mathbf{A}^{u}$ that plays precisely the same role in the time-dependent phonon-coupled Hamiltonian as the physical vector potential does for the photon coupling. Since the instantaneous eigenvalues of Eq.~(1) do not have physical meaning, one has to turn to Floquet-theory to relate these two different gauge fields to observable quantities. Indeed, performing Floquet-analysis of the time-dependent Hamiltonian Eq.~(1) gives a Floquet spectrum that is identical for the two bosonic excitations. They both only depend on $m$, the respective effective coupling strength.

It has been noted before that a lattice deformation of graphene can be described with an effective gauge field around the Dirac point\cite{Pisana:2007cl,Sasaki:2008hw,Pereira:2009iz} and the present work shows that this property also holds for dynamical phonons. By comparing the ARPES and Floquet spectra of the phonon-dressed electronic structure to the equivalent photon-dressed band structure in Figs.~3 and 4 we find indeed that the different dressing mechanisms result in the same spectra. Most importantly, the results reported in Fig.~4 show that such an effective gauge field creates the same non-equilibrium electronic structure as a physical vector potential. The coherent phonon modes we are considering here are $\Gamma$-phonons which implies that the gauge field that they induce is uniform across the crystal. This is in analogy to the dipole approximation used here to describe the photons via a uniform vector potential. However, the Floquet framework also holds for excitations with finite momentum transfers, i.e. lattice vibrations that induce spatial variations larger than the unit cell.    

%We note that the frozen-phonon approximation fails also here, because it is computed for a series of distorted lattices, each having a Dirac point slightly displaced from $K$~\cite{Pereira:2009iz} and thus resulting in an average with an open gap at $K$, which does not even qualitatively agree with the Floquet band structure. The Floquet band structure, instead, preserves the Dirac point as well as showing the complicated pattern of sideband crossings and avoided crossings, c.f. insets of Fig.~3. 

The implications of the equivalent behaviour are most striking when considering an excitation that breaks time-reversal symmetry. The Haldane model\cite{haldane_model_1988} predicts that the breaking of time-reversal symmetry in graphene leads to the emergence of  non-trivial topology, a Chern insulator. It has been shown that such a phase can indeed be created by circularly polarised lasers and the changes in the topological structure, Chern numbers, of the Dirac points have been identified with Floquet theory.\cite{Sentef:2015jp} Fig.~4(a) and Fig.~4(c) show the time-resolved ARPES and Floquet band structure for circularly polarized phonons and photons. Agreement for the two different excitations is observed underlining the fact that both dressed electronic structures originate form the same kind of Dirac Hamiltonian and implying that circularly polarized coherent phonons can induce the same kind of change in the topological order as photon fields. 

Having shown how Floquet analysis is required to understand the bosonic electron dressing at the Dirac point, we now consider what happens at the $\Gamma$ point, where the phonon-dressed electronic structure results in a strongly modified ARPES spectrum when compared to the equilibrium, c.f. Fig.~2(a). The apparent agreement of the cycle-averaged frozen phonon bands with the ARPES spectrum is misleading, because it wrongly displays a splitting of the degenerate $\sigma$-bands. In fact, as can be seen from the photo-electron spectrum and the underlying Floquet band structure, Fig.~2(a)~-~(b), there is no such splitting, but instead, the non-equilibrium bands are spread out in the usual sideband pattern, that is common for Floquet bands. Nevertheless, the phonon-quasiparticle dressing at this point is not a trivial process. The strong splitting of the cycle averaged frozen phonon bands at $\Gamma$ originates from a strong electron-phonon coupling. In the Floquet-phonon average however, this does not result in a splitting of bands, but instead the bands appear around the equilibrium position. The fractional occupations of the Floquet-sidebands, i.e. the time-averaged projections of the Floquet-state on the groundstate,\cite{Oka:2011ko} for these bands is spread across a region of $\sim10$ sidebands (c.f. Supplementary Fig.~S2(a)). That means that despite the appearance around the equilibrium position, there is a large interaction between the levels, corresponding to the large electron-phonon coupling. This results in the observed wide spread of spectral weight, thus mimicking the frozen phonon positions. Indeed, the agreement of the frozen phonon bands with the time-resolved ARPES spectrum is purely coincidental and a result of photo-electron matrix elements. In an optical absorption experiment the sidebands would appear as a series of satellite peaks, which the frozen phonon approximation would not be able to describe

\section{Conclusion}
We have demonstrated that the interacting electronic structure in presence of a coherent phonon gives rise to complex photo-electron spectra that reflect a non-equilibrium state of the system that can be fundamentally different from the groundstate. The underlying dynamical dressing process where the slow lattice motion creates additional energy levels for the electrons cannot be described by the frozen-phonon approach, despite the relatively large time-scale of the motion. Instead, the correct way to approach the electronic structure of such a system is Floquet-theory, where the time-average of the measurement process is performed while accounting for the quantum-mechanical nature of the dynamical interactions. The Floquet-phonon approach shows that the dynamical effect of a coherent phonon can be the same as the perturbation with an photon field, showing the fundamental equivalence between the two bosonic excitations. From such a treatment of graphene emerges a rich sideband structure that reveals the strong electron-phonon coupling at the Brillouin zone center and non-trivial topology of the non-equilibrium phase. 
%HH
The access to the non-equilibrium electronic structure provided by this approach can serve as a starting point for further first principles investigations of non-equilibrium effects of electron phonon coupling, for example giving access to non-equilibrium phonon bandstructures. 

Here we have considered the example case of graphene, to demonstrate the basic mechanism, but the present approach is designed to investigate other semi-metal systems, where the low energy splittings of phonon-dressing alters the Fermi-surface leading to complex changes in the material properties, for example by creating charge-density waves through finite momentum phonons or creating a non-equilibrium topological phase. Having shown the equivalence between photon and phonon excitations within this Floquet interpretation it can be expected that this general framework also applies to the coherent limit of other bosonic excitations such as magnons, plasmons or excitons.

\section{Acknowledgements}
We are grateful for illuminating discussions with I. Gierz, S. Aeschlimann, M. A. Sentef, and Th. Brumme. We acknowledge financial support from the European Research Council (ERC-2015-AdG-694097), Grupos Consolidados (IT578-13) and the European Union’s Horizon 2020 Research and Innovation program under Grant Agreements no. 676580 (NOMAD) and 646259 (MOSTOPHOS).

\section{Methods}
The evolution of the electronic structure under the effect of external fields was computed by propagating the Kohn-Sham (KS) equations in real-space and real time within TDDFT as implemented in the Octopus code.\cite{Strubbe:2015iz} 
We solved the KS equations in the local density approximation (LDA)~\cite{Perdew:1981dv} with semi-periodic boundary conditions. We used a simulation box of 60~$a_0$ along the non-periodic dimension and the primitive cell on the periodic dimensions with a grid spacing of 0.36~$a_0$. We modeled graphene with a lattice parameter of 4.651~$a_0$ and by sampling the Brillouin zone with a 12$\times$12 k-point grid. All calculations were performed using fully relativistic HGH pseudopotentials.\cite{Hartwigsen:1998dk}

The linearly polarized phonon mode was prepared by starting the time-evolution of the Kohn-Sham system from a distorted atomic configuration along the $C$-$C$ bond of 1\% of the lattice parameter. From this initial condition the lattice then evolved under Ehrenfest molecular dynamics as a stable oscillatory mode of 20.6~fs, corresponding to an energy of $\sim200$~meV. For the circular mode the time-dependence of the lattice was explicitly driven along the circular trajectory corresponding to a superposition of the TO and LO modes with a $\pi/2$ phase difference and with the same frequency as the liner mode. For the photon-excitations with photons shown in Fig.~1 and Fig.~3 the energy was chosen to be same as the phonon energy and the peak intensity was tuned to match the phonon sideband structure with $1.6\times10^9$ W/cm$^2$ for the linearly polarized photons, Fig.~1(b) and Fig.~3(c), and $3.6\times10^9$ W/cm$^2$ for the circularly polarized photons of Fig.~3(d). 

Time-resolved ARPES was calculated by recording the flux of the photoelectron current over a surface placed 30~$a_0$ away from the system with the t-SURFFP method.\cite{DeGiovannini:2016bb} To detect the phonon sideband structure a probe pulse of 50~eV, 80~fs length and a peak intensity of $10^9$ W/cm$^2$. For Fig.~2(a) a pulse length of 160~fs was used. 

The Floquet band structures of the driven system were computed using Floquet-TDDFT~\cite{Hubener:2017ht} by recording the time-dependent Kohn-Sham Hamiltonians during one cycle of the phonon mode and subsequently performing their Floquet analysis as described in the main text. We found that at least five sidebands were needed to converge the phonon-sideband structure, due to the strong interaction at the $\Gamma$ point. 

% \bibliography{FloquetPhonon}

\end{document}